\pdfoutput=1

\documentclass[11pt]{article}

\usepackage[final]{acl}

\usepackage{times}
\usepackage{latexsym}

\usepackage[T1]{fontenc}

\usepackage[utf8]{inputenc}

\usepackage{microtype}

\usepackage{inconsolata}
\usepackage{graphicx} 
\usepackage{algorithm}
\usepackage{algorithmic}
\usepackage{multirow}
\usepackage{subfigure}
\usepackage{amssymb}
\usepackage{multicol}
\usepackage{booktabs}
\usepackage{utfsym}

\usepackage{amsmath,amsfonts,bm}

\def\eqref#1{equation~\ref{#1}}

\def\1{\bm{1}}

\def\vzero{{\bm{0}}}

\def\vz{{\bm{z}}}

\def\mI{{\bm{I}}}

\DeclareMathAlphabet{\mathsfit}{\encodingdefault}{\sfdefault}{m}{sl}
\SetMathAlphabet{\mathsfit}{bold}{\encodingdefault}{\sfdefault}{bx}{n}

\def\gD{{\mathcal{D}}}

\def\gN{{\mathcal{N}}}

\def\gT{{\mathcal{T}}}

\def\sR{{\mathbb{R}}}

\def\sT{{\mathbb{T}}}

\DeclareMathOperator*{\argmin}{arg\,min}

\usepackage{graphicx}

\title{FlashAudio: Rectified Flows for Fast and High-fidelity Text-to-Audio Generation}

\author{%
  Huadai Liu$^{1,2}\thanks{Equal Contribution.}$, Jialei Wang$^{3}\footnotemark[1]$, Rongjie Huang$^{4}$, Yang Liu$^{3}$ \\ 
  \textbf{Heng Lu$^{2}$, Zhou Zhao$^{3}\footnotemark[2]$, Wei Xue$^{1}\thanks{Corresponding Author.}$} \\[1ex]
  $^{1}$Hong Kong University of Science and Technology
  \\ 
  $^{2}$Tongyi Lab, Alibaba Group \\ 
  $^{3}$Zhejiang University \\ 
  $^{4}$Fundamental AI Research (FAIR), Meta
}

\begin{document}
\maketitle
\begin{abstract}
  Recent advancements in latent diffusion models (LDMs) have markedly enhanced text-to-audio generation, yet their iterative sampling processes impose substantial computational demands, limiting practical deployment. While recent methods utilizing consistency-based distillation aim to achieve few-step or single-step inference, their one-step performance is constrained by curved trajectories, preventing them from surpassing traditional diffusion models. In this work, we introduce FlashAudio with rectified flows to learn straight flow for fast simulation. To alleviate the inefficient timesteps allocation and suboptimal distribution of noise, FlashAudio optimizes the time distribution of rectified flow with Bifocal Samplers and proposes immiscible flow to minimize the total distance of data-noise pairs in a batch vias assignment. Furthermore, to address the amplified accumulation error caused by the classifier-free guidance (CFG), we propose Anchored Optimization, which refines the guidance scale by anchoring it to a reference trajectory. Experimental results on text-to-audio generation demonstrate that FlashAudio's one-step generation performance surpasses the diffusion-based models with hundreds of sampling steps on audio quality and enables a sampling speed of 400x faster than real-time on a single NVIDIA 4090Ti GPU. Code will be available at~\url{https://github.com/liuhuadai/FlashAudio}.
  ~\footnote[1]{Audio Samples are available at \url{https://FlashAudio-TTA.github.io/.}}

\end{abstract}

\section{Introduction}
\begin{figure}[t]
\centering
    \includegraphics[width=0.495\textwidth]{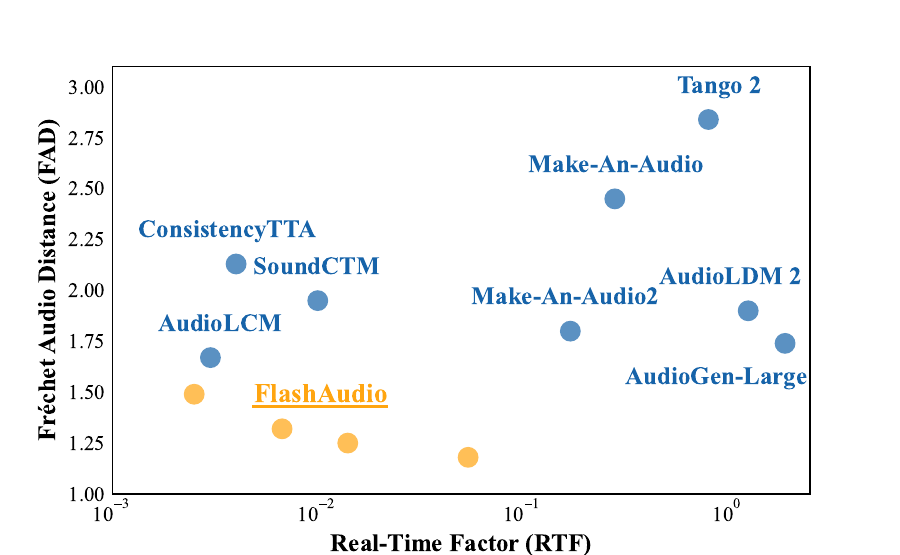}
    \caption{FlashAudio is a fast and high-fidelity text-to-audio generation model with rectified flow. It has a remarkable speed with 400x faster than real-time, on a single NVIDIA 4090Ti GPU.}
    \vspace{-5mm}
\end{figure}
Diffusion models~\citep{ho2020denoising,ddpm,song2021scorebasedgenerativemodelingstochastic} have demonstrated significant capabilities in modeling various modalities, including images~\citep{rombach2022high,saharia2022photorealistic}, audio~\citep{liu2023audioldm,huang2023make}, and video~\citep*{ho2022imagenvideohighdefinition,gupta2023photorealisticvideogenerationdiffusion}.They have become the de-facto approach for generating high-fidelity audio from natural language inputs due to their impressive generalization capabilities.
However, their iterative nature and the associated computational costs, along with prolonged sampling times during inference, limit their application in real-time scenarios~\citep{song2023consistency}.
Recent advances~\citep{salimans2022progressivedistillationfastsampling,kim2024consistencytrajectorymodelslearning} have thus concentrated on distillation models that estimate the integral along the Probability Flow (PF) ODE sample trajectory, effectively reducing the computational load associated with numerical solvers.
For instance, AudioLCM~\citep{liu2024audiolcmtexttoaudiogenerationlatent} utilizes guided consistency distillation with multi-step ODE solvers, achieving two-step generation performance on par with diffusion models. In contrast, some studies adopt one-step generation~\citep{saito2024soundctm,bai2024consistencyttaacceleratingdiffusionbasedtexttoaudio} but struggle to align with latent diffusion models.

While mapping any point in the trajectory into initial point leads to efficient training,  it introduces the challenge of estimating data points along curved trajectories. This choice can significantly impact sampling, as demonstrated by~\citet{kim2024consistencytrajectorymodelslearning}. 
For instance, a forward process that fails to accurately map from noise to data may lead to a discrepancy between the training and test distributions, resulting in artifacts such as noisy audio samples.
Although curved paths require many integration steps to simulate the process, a straight path can be simulated in a single step, reducing the risk of error accumulation~\citep{esser2024scaling}. Since each step corresponds to an evaluation of the neural network, this directly influences the sampling speed.

Rectified Flow (RF)\citep{liu2022flow} is a novel flow-based generative model that linearly transfers the source distribution to the target distribution. This property of linear transfer facilitates simulation-free training and faster sampling at inference time. Recently, RF-based methods have garnered significant attention in image synthesis\citep{liu2023instaflow,esser2024scaling} and text-to-speech applications~\citep{guan2024reflowttsrectifiedflowmodel}. For instance, InstaFlow~\citep{liu2023instaflow} generates high-quality images in just one step. While this model class offers superior theoretical properties and has proven effective in both image and speech domains, it has not yet been explored for text-to-audio generation.


In this paper, we introduce FlashAudio, a text-to-audio generation model using rectified flows to learn straight flows for fast and high-fidelity output. Traditional RFs uniformly distribute time steps, often resulting in suboptimal performance, as easier tasks consume resources that could be better allocated to more challenging ones. To overcome this, we propose the Bifocal Samplers, which reallocates time steps to focus on more difficult aspects of the task, improving both model efficiency and stability. Additionally, we introduce immiscible flow to minimize the total distance of data-noise pairs in a batch vias assignment.  Finally, to mitigate the amplified accumulation error caused by classifier-free guidance (CFG), we propose Anchored Optimization, which refines the guidance scale by anchoring it to a reference trajectory at $\omega = 1$, enhancing both the audio quality and text-audio alignment with larger guidance scales.

Experimental results on text-to-audio generation demonstrate that FlashAudio achieves state-of-the-art performance in both objective and subjective metrics with significantly reduced inference time. FlashAudio also outperforms strong diffusion-based baselines and consistency-based models in both few-step and one-step generation settings. Our extensive preliminary analysis and ablation studies show that each design in FlashAudio is effective. The main contributions of this study are summarized below:

\vspace{-2mm}
\begin{itemize}
    \item We propose FlashAudio with rectified flows for fast and high-fidelity text-to-audio generation, which is the first work in TTA to learn straight flows for fast simulation.
    \vspace{-1mm}
    \item FlashAudio includes bifocal samplers and immiscible flow techniques for efficient and stable training of rectified flow models.
    \vspace{-1mm}
    \item FlashAudio introduces anchored optimization to refine the guidance scale by anchoring it to a reference trajectory.
    \vspace{-1mm}
    \item Experimental results demonstrate that FlashAudio achieves state-of-the-art performance with much less inference time in both multi-step and one-step generation. This makes generative models practically applicable for text-to-audio generation deployment.
\end{itemize}

\section{Preliminaries}
In this section, we briefly introduce the theory of flow matching and rectified flow.

\subsection{Flow Matching}
Flow matching (FM)~\citep{chen2018neural} is a generative model for training objective to regress onto a target vector field that generates a desired probability path. Let $\sR^d$ denote the data space with data point $x_0$, flow matching aims to learn
vector field $v_\theta(t, x, c): [0,1]\times \sR^d \mapsto \sR^d$, such that the solution of the ordinary differential equation (ODE) can transfer standard Gaussian distribution $z_0 \sim \pi_0 $ to latent audio distribution
$z_1 \sim \pi_1$:
\begin{equation}
    \frac{\phi_t(z)}{dt} = v_\theta(\phi_t(z),t)
    \label{eq:1}
\end{equation}
where we model the vector field $v_t$ with a neural network $v_\theta(z,t)$. Given the ground truth vector field $u(z,t)$ that generates probability path $\pi_t$ under the two marginal constraints
that $\pi_{t=0} = \pi_0$ and $\pi_{t=1} = \pi_1$, the vector field is learned by minimizing a simple mean square
objective (MSE):
\begin{equation}
    \min_v \mathbb{E}_{(z_0,z_1)\sim \gamma}\| v_\theta(z_t,t) - u(z_t,t)\|^2
    \label{eq:2}
\end{equation}
where the $\gamma$ is any coupling of $(\pi_0,\pi_1)$.

However, it is computationally intractable to find such $u$ and $\pi_t$.  Instead of directly optimizing Eq.~\ref{eq:2}, Conditional Flow Matching (CFM)~\citep{lipman2022flow} regress $v_\theta(z,t,c)$ on
the conditional vector field $u(t, z_t|z_1)$ and probability path $\pi_t(z_t|z_1)$:
\begin{equation}
    \min_v \mathbb{E}_{(z_0,z_1)\sim \gamma}\| v_\theta(z_t,t) - u(z_t|z_1,t)\|^2
    \label{eq:4}
\end{equation}


\subsection{Rectified Flow}
The rectified flow~\citep{liu2022flow} is an ODE model that transport distribution $\pi_0$ to $\pi_1$ by following straight line
paths as much as possible. 
In rectified flow, the drift force
$v$ is set to drive the flow to follow the direction $(z_1 - z_0)$ of the linear path pointing from $z_0$ to $z_1$
as much as possible:
\begin{equation}
    \min_v \mathbb{E}_{(z_0,z_1)\sim \gamma}\| (z_1 - z_0) - v_\theta(z_t,t) \|^2
    \label{eq:5}
\end{equation}
where $z_t$ is the linear interpolation of $z_0$ and $z_1$ and $z_t = (1-t)z_0 + tz_1$.
\paragraph*{\textbf{Reflow}} Reflow is an iterative procedure to straighten the trajectories without modifying the marginal distributions, hence allowing fast simulation at inference time. In text-to-audio generation, the reflow objective
is as follows:
\begin{equation}
    v_{k+1} = \argmin_v \mathbb{E}_{z_0\sim \pi_0}\| (z_1 - z_0) - v_\theta(z_t,t,c)\|^2 
    \label{eq:6}       
\end{equation}
where $c$ is the text embeddings and $z_1 = \text{ODE}[v_k](z_0 | c)$ and $v_{k+1}$ is optimized using the objective equation~\ref{eq:5}, but with $(Z_0,Z_1)$ pairs constructed from the previous $\text{ODE}[v_k]$.

\section{FlashAudio}
This section introduces FlashAudio, a novel rectified flow framework designed for fast and high-fidelity text-to-audio generation. We begin by discussing the motivation behind each design choice in FlashAudio. Subsequently, we detail the selection process for a powerful pre-trained CFM model to initialize the 1-rectified flow model. Next, we elaborate on our advanced training techniques, including bifocal samplers and the immiscible flow method. This is followed by an introduction to reflow and anchored optimization for better few-step performance. Finally, we employ the distillation for one-step generation and display the training, inference procedures and algorithm employed in FlashAudio

\subsection{Motivation}

Diffusion models~\citep{song2019generative,song2020denoising} have made notable advancements in domains such as image and audio generation. However, the iterative nature of current latent diffusion text-to-audio models requires extensive computational resources, leading to slow inference times and limited real-time applicability~\citep{luo2023lcm}. This constraint impedes the practical deployment of these models in real-world scenarios. Recent methods~\citep{liu2024audiolcmtexttoaudiogenerationlatent} have employed consistency distillation to enhance inference speed by mapping points on curved trajectories with initial points. Nonetheless, curved trajectories are prone to greater error accumulation compared to straight paths, which restricts their performance relative to diffusion models in few-step settings.

In contrast, rectified models~\citep{liu2022flow} offer a novel generative approach by mapping data and noise along straight paths, facilitating efficient few-step and one-step generation. Despite their promising theoretical benefits, rectified models have been under-explored in the context of text-to-audio generation. To bridge this gap, we introduce advanced training techniques for rectified flow models, such as immiscrible flow and bifocal samplers, to ensure efficient and stable training. Furthermore, to address the error amplification caused by Classifier-Free Guidance (CFG)~\citep{classifier-free} during reflow and couplings generation, we propose anchored optimization to enhance sample accuracy for subsequent reflow procedures.

\subsection{Initialization from Pre-trained Flow Matching Models}
As a blossoming class of generative models, flow matching models~\citep{lipman2022flow} have demonstrated their powerful generation abilities across image~\citep{gat2024discreteflowmatching} and audio~\citep{vyas2023audioboxunifiedaudiogeneration}, with less sampling steps and more high-quality generated samples. Compared to training rectified flow models from scratch, initializing the neural network with flow matching model benefits in inheriting powerful capabilities and speeding up convergence. 
The experimental results in Section~\ref{4.4} demonstrate the advantage of adopting the pre-trained models.

\subsection{Advanced Techniques for RF Models}
\subsubsection{Bifocal Samplers for Timesteps}
\label{Bifocal}
In the training of rectified flow models, the common approach is to sample timesteps $t$ uniformly across the interval $[0,1]$, i.e., $t \sim \text{Uniform}(0,1)$. However, the choice of sampling distribution significantly influences the effectiveness of the training process. Ideally, more computational resources should be allocated to the timesteps that are more challenging for the model, rather than distributing effort uniformly.

During 1-rectified flow training, the model learns to approximate the dataset's average when $t=1$ and the noise average when $t=0$. This makes the intermediate timesteps within $[0,1]$ particularly challenging. A practical solution to this issue is to modify the time distribution from the standard uniform distribution to one that prioritizes intermediate timesteps. Inspired by \citet{esser2024scaling}, we adopt the logit-normal distribution~\citep{atchison1980logistic} $p(t)$ for this purpose:
\begin{multline}
    p_1(t; \mu, \sigma^2) = \frac{1}{\sqrt{2 \pi \sigma^2}} \frac{1}{t(1-t)} \\
    \times \exp \left( -\frac{1}{2 \sigma^2} \left[ \text{logit}(t) - \mu \right]^2 \right),
\end{multline}

where the logit function is defined as:
\begin{equation}
\text{logit}(y) = \log \left(\frac{y}{1 - y}\right).
\end{equation}
In this formulation, \(\mu\) and \(\sigma^2\) denote the mean and variance in the logit space, respectively. Figure~\ref{fig:sampler} provides a visual representation of the logit-normal distribution.

For the reflow process, which is applied only once in FlashAudio, the 2-rectified flow model learns to directly predict data from noise at $t=1$ and noise from data at $t=0$. So it is non-trivial to assign greater emphasis to the timesteps near the noise and data. To tackle this, we employ a mixture of exponential distributions (Mix-Exp) defined as:
\begin{equation}
    p_2(t) \propto \exp(at) + \exp(-at),
\end{equation}
where $a$ serves as the scale parameter. The distribution of Mix-Exp is depicted in Figure~\ref{fig:sampler}.

\subsubsection{Immiscible Flow Mechanism}
The phenomenon of particles becoming tightly jumbled together during diffusion, making them difficult to separate individually, is akin to the challenges observed in rectified flow models. When particles are rendered immiscible, they maintain a similar overall distribution but remain distinctly identifiable. This concept can be translated into rectified flow models where each audio sample can map to any point in the noise space, and vice versa, making it challenging for the model to differentiate during the reverse process.

To simulate this immiscible phenomenon, we propose a method where each noise point is assigned to a limited set of audio samples, reducing confusion for the rectified flow model. Despite this, the noise space must remain strictly Gaussian to ensure efficient sampling. Drawing inspiration from \citet{li2024immisciblediffusionacceleratingdiffusion}, we introduce the concept of Immiscible Flow. This approach involves assigning batches of noise samples to corresponding batches of audio samples based on their proximity in a shared space, minimizing the total distance between data-noise pairs during training.

The noise remains Gaussian after assignment, with each noise point being associated with closer audio samples, analogous to the immiscible phenomenon. This significantly alleviates the challenges associated with the simulation process. Practically, this assignment can be efficiently executed using the Hungarian algorithm, which can be implemented with Scipy~\citep{virtanen2020scipy} as follows:
\begin{equation}
    \begin{aligned}
        \mathbf{M} &= \text{linear\_sum\_assignment}(\text{dist}(z_1, z_0)) \\
        z_0^{\prime} &= z_0[\mathbf{M}] \\
        z_t &= (1-t) z_0^{\prime} + t z_1
    \end{aligned}
\end{equation}

\subsection{Reflow and Anchored Optimization}
\subsubsection{Straightening Flows via Reflow}
To achieve optimal performance in few-step generative tasks, it is essential to ensure that the flow trajectories are as straight as possible. Training a rectified flow model just once is often insufficient for constructing adequately straight transport paths. 
Therefore, an additional reflow process is applied using the training objective defined in Eq.~\ref{eq:6} to straighten the transport paths. As depicted in Figure~\ref{fig:straight}, the 2-rectified flows achieve near-straight trajectories, effectively eliminating the need for further rectification (e.g., 3-rectified flow).

\subsubsection{Anchored Optimization}\label{3.4.2}

Classifier-Free Guidance (CFG)~\citep{classifier-free} plays a critical role in generating audios that align with text prompts. During sampling, \citet{liu2023instaflow} introduce a guided velocity field in the context of text-conditioned rectified flows using the formula:
\vspace{-2mm}
\begin{equation}
    v_{cfg}(z_t, t,c) = \omega v(z_t, t, c) + (1 - \omega) v(z_t, t, \varnothing)
\end{equation}
where $\omega$ is the guidance scale. 

In practice, the multiple regression of the velocity field $v_\theta$ and the application of sampling function in generating $\hat{z}_1$ inevitably incorparate a slight error in every step. These errors are further amplified when a larger guidance scale $\omega > 1$ is used, which is likely to perturb the original marginal distribution. To alleviate these errors, InstalFlow employs a much smaller guidance scale ($\omega = 1.5$) in the final stage while their initial scale is 5. 
While using $\omega = 1$ can roughly preserve the original marginal distribution, the lack of CFG makes it struggle to balance the audio quality and alignment with text prompts.

In $\hat{z}_1$ generation stage, inspired by~\citet{mokady2023null}, we use the initial simulation with $\omega = 1$ as a pivot trajectory and optimize around it with a guidance scale $\omega > 1$ during all paired generation stages. Direct optimization of textual embeddings often leads to non-interpretable representations, as the optimized tokens may not correspond to actual words. Instead, we leverage CFG's core characteristic: the outcome is strongly influenced by the unconditional prediction. Therefore, we focus on optimizing only the unconditional embedding $\varnothing$ while keeping the model parameters and textual embeddings fixed.

In practice, simulation with $\omega = 1$ generates a sequence \( z_0, z^*_{1/T}, \ldots, z^*_1 \). We initialize with the same Gaussian noise \( z_0 \) and perform the following optimization with guidance scale \( \omega > 1 \) for timesteps \( t = 1/T, 2/T, \ldots, 1 \), iterating \( T \) times at each step:

\begin{equation}
    \min_{\varnothing_t} \| z^*_{t-1} - z_{t-1} \|^2
\end{equation}

Here, \( z_t \) represents the intermediate result of the guided sampling. Early stopping is employed to minimize computational time.

In reflow training, we use $v_{cfg}$ as the target of regression, the reflow objective can be written as:
\begin{equation}
    \min_v \mathbb{E}_{(z_0,z_1)\sim \gamma}\| (z_1 - z_0) - v_{cfg}(z_t,t) \|^2
    \label{eq:12}
\end{equation}
\subsection{Distillation for One-step Generation}
As shown in Figure~\ref{fig:straight}, the trajectories of the 2-rectified flow model become nearly straight after a single reflow step. With these approximately straight ODEs, a promising approach to enhance the performance of one-step generation is through one-step distillation:

\begin{equation}
    \hat{v}_{k} = \argmin_v \mathbb{E}_{z_0 \sim \pi_0} \| \text{ODE}[v_k] - \left(z_0 + v_{cfg}(z_0,c)\right)\|^2 
    \label{eq:13}
\end{equation}

In this process, we learn a single Euler step $z + v(z|\sT)$ to compress the mapping from $z_0$ to $\text{ODE}[v_k](z_0|\gT)$, effectively simplifying and accelerating the generation process.

\subsection{Training and Inference Procedures}
The training algorithm for FlashAudio, incorporating rectified flows, is detailed in Algorithm~\ref{alg:1}.

\begin{algorithm}[ht]
    \centering
    \caption{Training FlashAudio with Rectified Flows}\label{alg:training}
    \begin{algorithmic}[1]
        \STATE \textbf{Input}: Initialize the velocity field network with pre-trained flow matching; a dataset of text prompts $\gD_\gT$; encoding training data into latent space $\gD_z$.
        \STATE Sample $\vz_{0} \sim \gN(\vzero,\mI)$ and $\vz_{1} \sim \gD_z$.
        \STATE Re-weight timestep $t$ using the \textbf{logit-normal distribution} and assign $z_0$ via \textbf{immiscible flow}.
        \STATE Train the 1-rectified flow $v_1$ by minimizing the objective in Equation~\ref{eq:5}.
        \STATE Generate the couplings $[z^{'}_0,z^{'}_1]$ for reflow with \textbf{anchored optimization}.
        \STATE Re-weight timestep $t$ using the \textbf{Mix-Exp distribution} and assign $z_0$ via \textbf{immiscible flow}.
        \STATE Train the 2-rectified flow $v_2$ by minimizing the objective in Equation~\ref{eq:12}.
        \STATE Generate the couplings $[z^{''}_0,z^{''}_1]$ with \textbf{anchored optimization} for one-step distillation.
        \STATE Train the final model $\hat{v}_2$ by minimizing the objective in Equation~\ref{eq:13}.
    \end{algorithmic}
    \label{alg:1}
\end{algorithm}
\vspace{-2mm}
\subsubsection{Training Procedure}
The training process begins by training the 1-rectified flow models using a pre-trained flow matching model, followed by the generation of reflow couplings. Our experiments indicate that the 1-rectified model performs exceptionally well in the CFG setting, whereas the 2-rectified model with $\omega > 1$ fails to achieve comparable performance with $\omega = 1$. For detailed experimental analysis, please refer to Section~\ref{4.2.3}. Therefore, anchored optimization is conducted when generating couplings $[z^{'}_0, z^{'}_1]$ and $[z^{''}_0, z^{''}_1]$. All rectified models utilize Mean Squared Error (MSE) loss as the distance function.

\subsubsection{Inference Procedure}
For inference, we adopt the Euler Solver for fast simulation, which is governed by the equation:

\begin{equation}
    z_{t+\Delta t} = z_t + \Delta t \, v(z_t, t | \gT) 
\end{equation}

Here, the simulation is performed with a step size of $\Delta t = 1/T$, completing the process in $T$ steps. In the few-step setting, we utilize the $v_2$ model for sampling. In the one-step setting, we employ the $\hat{v}_2$ model to generate audio latents. These latents are then transformed into mel-spectrograms, which are subsequently used to generate waveforms using a pre-trained vocoder.

\section{Experiments}
\subsection{Experimental setup}
\subsubsection{Dataset} \label{data}
For text-to-sound generation, we use the training split of Audiocaps dataset~\citep{kim2019audiocaps} to train all our models. For text-to-music generation, we exclusively employ the LP-Musicaps~\citep{doh2023lpmusiccaps} dataset for FlashAudio training endeavors. 
\subsubsection{Model configurations}
Initialized model is originally trained with objective equation~\ref{eq:4}, we utilize its pre-trained VAE, a continuous 1D-convolution-based network. This VAE is used to compress the mel-spectrogram into a 20-channel latent representation with a temporal axis downsampling rate of 2. Training 1-rectified flow involves 40,000 iterations on 4 NVIDIA 4090Ti GPU, with a batch size of 16 per GPU. We use the AdamW optimizer with a learning rate of 4.8e-5. Our choice for the vocoder is BigVGAN-v2~\cite{leeBigVGANUniversalNeural2023}, which is known for its universal applicability to different scenarios. We train the vocoder on the AudioSet dataset to ensure robust performance. More details on the model configuration can be found in Appendix~\ref{A.2}.
\begin{figure}[t]
    \centering
  \includegraphics[scale=.47]{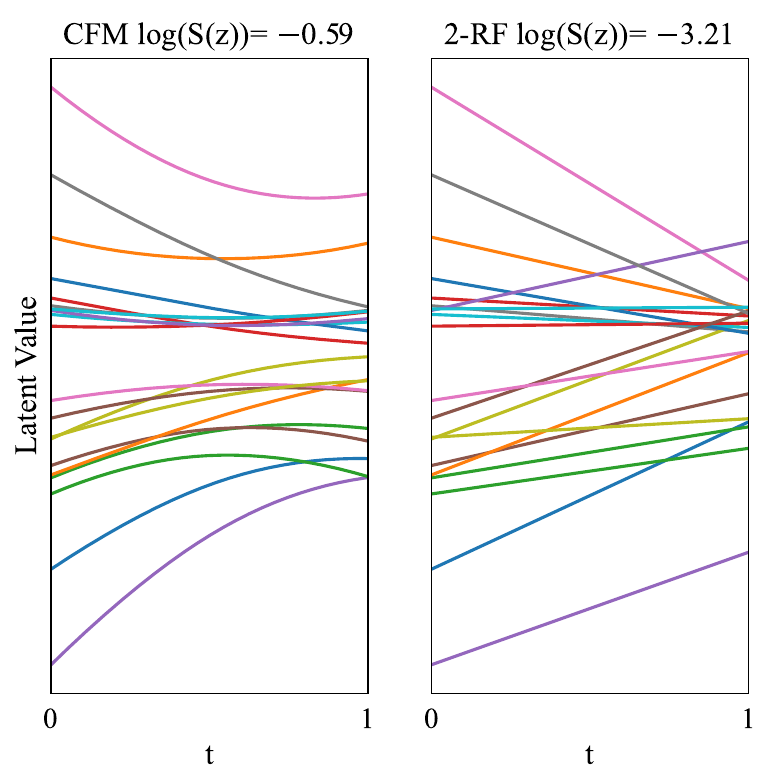}
  \vspace{-1mm}
  \caption{The straightness and simulation figure of flow matching and 2-rectified flow (2-RF) models. The top of each subfigure has its corresponding value of $\log(S(z))$}.
  \label{fig:straight}
  \vspace{-6mm}
\end{figure}
\subsubsection{Evaluation Metrics}
\label{metrics}
Our models conduct a comprehensive evaluation~\citep{cui2021emovie} using both objective and subjective metrics to measure audio quality, text-audio alignment fidelity, and inference speed. Objective assessment includes Kullback-Leibler (KL) divergence, Frechet audio distance (FAD), and CLAP score to quantify audio quality. The Real-time Factor (RTF) is also introduced to measure the system's efficiency in generating audio for real-time applications. RTF is the ratio between the total time taken by the audio system to synthesize an audio sample and the duration of the audio.
In terms of subjective evaluation, we conduct crowd-sourced human assessments employing the Mean Opinion Score (MOS) to evaluate both audio quality (MOS-Q) and text-audio alignment faithfulness (MOS-F). Detailed information regarding the evaluation procedure can be accessed in Appendix~\ref{A.4}.

\subsection{Preliminary Analyses}
In this section, we analyze 2-rectified flows with a range of few-step settings (from 16 to 1) and investigate the influence of CFG on the training of rectified flows. Additionally, we explore the effects of reflow on straightening flows and clarify why reflow is applied only once.

\subsubsection{Straighen Effects of Reflow}
We evaluate the straightness of the 2-rectified flow model and analyze the effectiveness of the reflow procedure in achieving straighter flows. Following the methodology of InstalFlow, we quantify the straightness by measuring the deviation of the velocity along the trajectory. Specifically, we use the metric:
\begin{equation}
    S(z) = \int_{0}^{1} \mathbb{E}\left[\| (z_1 - z_0) - v(z_t, t) \|^2\right] \, dt,
\end{equation}
where \( S(z) \) represents the mean squared deviation of the velocity from the desired trajectory. We compare the trajectories of flow matching and 2-rectified flow models through simulation. As illustrated in Figure~\ref{fig:straight}, reflow significantly reduces the estimated \( S(z) \), demonstrating its effectiveness in straightening the flow trajectories. Furthermore, the simulation results show that the 2-rectified model achieves nearly straight trajectories.

\subsubsection{Straight Flows Yield Fast Generation}
\begin{figure}[htbp]
    \centering
  \includegraphics[width=.49\textwidth]{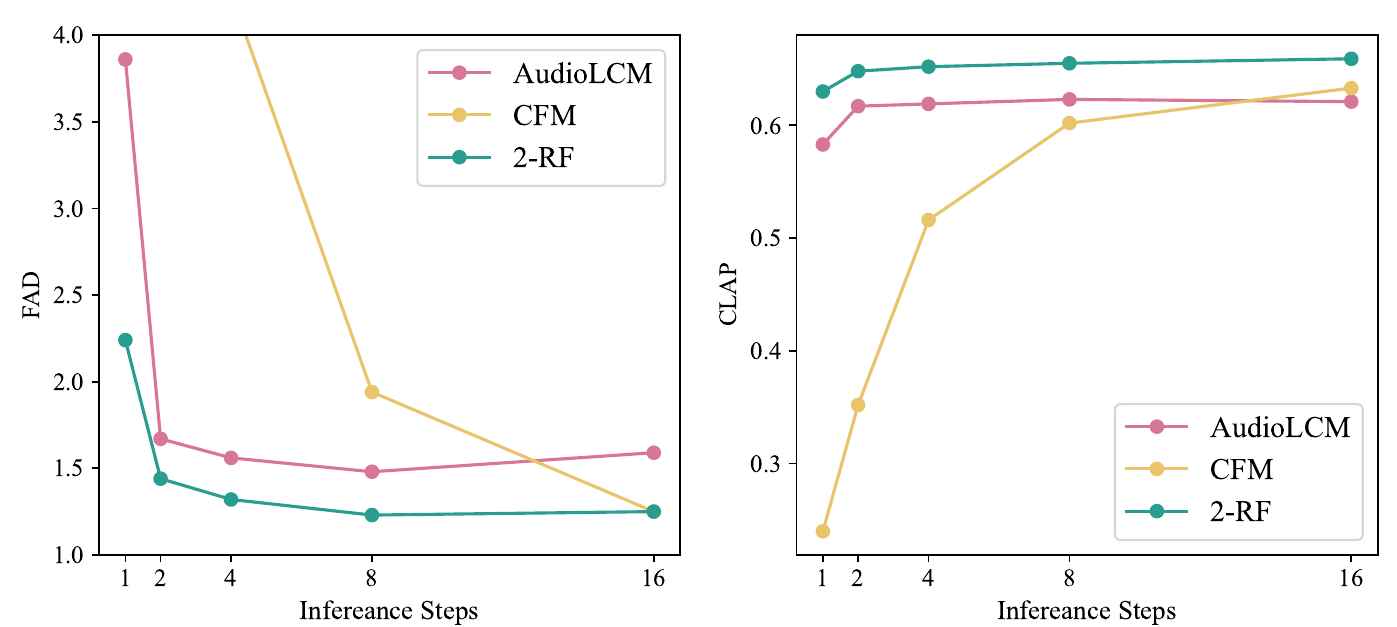}
  \caption{Comparisons with AudioLCM and conditional flow matching in few-step setting, measured by fad and CLAP score.}
  \label{fig:few}
  \vspace{-4mm}
\end{figure}

In both diffusion and flow matching models, selecting the sampling step \( T \) involves a trade-off between computational cost and accuracy: a larger \( T \) provides a better approximation of the ODE but increases computational expense, while a smaller \( T \) may struggle to maintain accuracy. For efficient simulation, it is essential to learn ODEs that can be accurately and rapidly simulated with a small \( T \). To address this, we compared our nearly straight 2-RF models with curved baselines such as flow matching and consistency models in a few-step setting. The results, presented in Figure~\ref{fig:few}, lead to several key observations: (1) Compared to curved flow matching models, our proposed 2-RF models consistently deliver superior performance across all sampling steps, from 16 to 1. This finding underscores the importance of straightening the ODE trajectories to enhance performance in fast generation tasks. (2) Our 2-RF models also outperform AudioLCM, particularly in the one-step setting, with a lower FAD (2.24 vs. 3.86) and a higher CLAP score (0.63 vs. 0.583). This demonstrates the superiority of our rectified flow approach in few-step generation scenarios.

\subsubsection{Impact of CFG on Audio Quality}\label{4.2.3}
Rectified Flows utilize reflow to straighten trajectories while maintaining the marginal distribution. However, the deployment of CFG is likely to perturb the distribution. We conduct experiments on rectified flow to explore the impact of CFG in RF and the effect of anchored optimization.
The key observations are as follows: (1) As the guidance scale increases, there is a notable decline in audio quality for original RF models. This supports our assumption in Section~\ref{3.4.2}. (2) After adding anchored optimization, it leads to a significant improvement in audio quality and text alignment. Specifically, the FAD score decreases from 1.43 to 1.26 and the CLAP score increases from 0.639 to 0.652. This demonstrates the effectiveness of our proposed anchored optimization.

\begin{figure}[htbp]
    \centering
  \includegraphics[width=.49\textwidth]{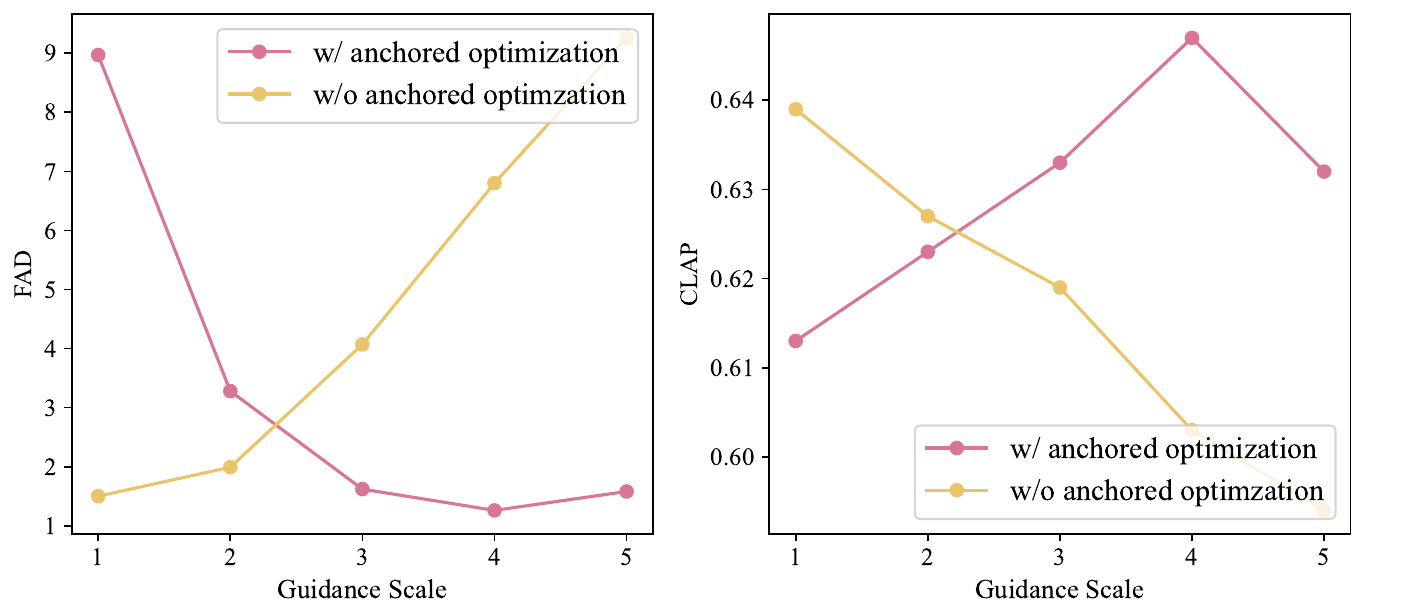}
  \vspace{-1mm}
  \caption{Comparison between with and without Anchored Optimization varying different guidance scale.}
  \label{fig:2-rf}
  \vspace{-4mm}
\end{figure}

\subsection{Performance on Text-to-Sound Generation}
\begin{table*}[htbp]
\centering
  \small
  \vspace{-3mm}
  \begin{tabular}{l|c|cccc|cc}
  \toprule
  \multirow{2}{*}{\bfseries Model} & \multirow{2}{*}{\bfseries NFE} & \multicolumn{4}{c|}{\bfseries Objective Metrics} & \multicolumn{2}{c}{\bfseries Subjective Metrics} \\
                         & & \bfseries FAD ($\downarrow$) & \bfseries KL ($\downarrow$) &\bfseries CLAP ($\uparrow$) & \bfseries RTF ($\downarrow$) &\bfseries MOS-Q($\uparrow$)   & \bfseries MOS-F($\uparrow$)   \\
  \midrule
  GT                                              & / & / & / & 0.670 & / & 87.90 & 85.48 \\
  \midrule            
  AudioGen-Large                                  & / & 1.74 & 1.43 & 0.601 & 1.890 & / & / \\
  Make-An-Audio                                  & 100 & 2.45 & 1.59 & 0.616 & 0.280 & 69.79 & 66.19 \\
  AudioLDM 2                                 & 100 & 1.90 & 1.48 & 0.622 & 1.250 & 73.38 & 71.22 \\
  Make-An-Audio 2                                  & 100 & 1.80 & 1.32 & 0.645 & 0.170 & 75.56 & 73.14 \\
  Tango 2                                  & 100 & 2.84 & \bfseries 1.20 & \bfseries 0.680 & 0.800 & 73.46 & 72.08 \\
  CFM                                  & 24 & 1.22 & 1.34 &  0.640 & 0.054 & 76.32 & 74.75 \\
  \textbf{FlashAudio ($\mathcal{R} $ \usym{2713} $\mathcal{D} $ \usym{2717})}                       & 24 & \bfseries 1.18 &  1.28 &  0.658 & 0.054 & \bfseries 78.86 & \bfseries 76.98 \\
  \textbf{FlashAudio ($\mathcal{R} $ \usym{2713} $\mathcal{D} $ \usym{2717})}                       & 4 & 1.26 & 1.30 &  0.652 & \bfseries 0.014 & 78.23 & 76.14 \\
  \midrule
  AudioLCM &    2 &  1.67 &  1.37 & 0.617 & 0.003 & 76.48 &  73.92 \\
  ConsistencyTTA & 1 & 2.13 & 1.33 & 0.655 & 0.004 & 73.19 & 70.08 \\
  SoundCTM & 1 & 1.95 & 1.36 & \bfseries 0.656 & 0.01 & 73.87 & 71.16 \\
  \bfseries FlashAudio ($\mathcal{R} $ \usym{2713} $\mathcal{D} $ \usym{2717}) & 1 & 2.24 & 1.40 &  0.630 & 0.0025 & 72.67 & 70.84 \\
  \bfseries FlashAudio ($\mathcal{R} $ \usym{2713} $\mathcal{D} $ \usym{2713}) & 1 & \bfseries 1.49 & \bfseries 1.32 & 0.648 & \bfseries 0.0025 & \bfseries 77.56 & \bfseries 75.43 \\
  \bottomrule
  \end{tabular}
  \vspace{1mm}
  \caption{The audio quality and sampling speed comparisons. The evaluation is conducted on a server with 1 NVIDIA 4090Ti GPU and batch size 1. NFE (number of function evaluations) measures the computational cost, which refers to the total number of times the denoiser function is evaluated during the generation process. $\mathcal{R}$ denotes reflow and $\mathcal{D}$ denotes one-step distillation. }
  \vspace{-5mm}
  \label{table:audio}
\end{table*} 

We conduct a comparative analysis of the quality of generated audio samples and inference latency across various systems, including GT (i.e., ground-truth audio), AudioGen~\citep{kreuk2023audiogen}, Make-An-Audio~\citep{huang2023makeanaudio}, AudioLDM 2~\citep{liu2023audioldm}, TANGO 2~\citep{majumder2024tango2aligningdiffusionbased}, Make-An-Audio 2~\citep{huang2023makeanaudio2}, ConsistencyTTA~\citep{bai2024consistencyttaacceleratingdiffusionbasedtexttoaudio}, SoundCTM~\citep{saito2024soundctm}, AudioLCM~\citep{liu2024audiolcmtexttoaudiogenerationlatent}, and our constructed CFM. utilizing the published models as per the respective paper. The evaluations are conducted using the AudioCaps test set and then calculate the objective and subjective metrics. The results are compiled and presented in Table~\ref{table:audio}. From these findings, we draw the following conclusion:

\textbf{Audio Quality} In terms of audio quality, we first compare FlashAudio's 2-RF models with our constructed CFM and other baselines. With the same number of steps as CFM, FlashAudio not only outperforms CFM but also significantly surpasses other baselines, achieving a notably lower FAD score of 1.18. In a few-step setting with four steps, FlashAudio remains competitive with CFM and markedly outperforms diffusion-based models and AudioLCM. Further evaluation of FlashAudio’s one-step generation performance, enhanced by additional distillation, shows substantial improvements in both objective and subjective metrics. FlashAudio excels over consistency-based models in nearly all metrics, aside from a slightly lower CLAP score, and also achieves superior audio quality compared to diffusion-based models—a notable advancement that consistency-based models have yet to achieve. These observations suggest that FlashAudio possesses the capability to generate high-fidelity audio with fast simulation.

\textbf{Sampling Speed}
FlashAudio surpasses diffusion models and consistency-based models with an exceptional RTF of 0.0025, demonstrating superior speed while maintaining competitive audio quality. This translates to a remarkable speed, approximately 400x faster than real-time, on a single NVIDIA 4090Ti GPU. This impressive reduction in inference time establishes FlashAudio as a leading solution in efficient TTA generation.
\label{4.4}
\subsection{Ablation Studies}
In this section, we conduct ablation studies to validate the effectiveness of initialization from CFM, our proposed bifocal samplers, and immiscible flow. We attach more exploratory experimental results to Appendix~\ref{A.3}.

  The results are presented in Table~\ref{table:ablation}, and we have the following observations: (1) 
  The significant improvement of all metrics from ID 1 to 0 demonstrates the effectiveness of initialization from pre-trained condition flow matching.
 (2) Although the adoption of Logit-Norm in the training of 1-rectified flow slightly causes the decrease of FAD score, the improvement of KL and CLAP validates the positive impact of re-weighting timesteps in middle timesteps, with the KL from 1.27 to 1.25 and CLAP score from 0.649 to 0.659. And ID 4 outperforms ID 5 across all objective metrics, which shows the necessity to concentrate on the boundary of tiemsteps between [0,1]. These two sets of results demonstrate the effectiveness of our proposed bifocal samplers for improving the training of rf. (3) The degradation from ID 0 to ID 3 and from ID 4 to ID 6 proves that the immiscible flow mechanism plays an important role in RFs training.

 \begin{table}[htbp]
  \centering
    \small
    \begin{tabular}{c|c|ccc}
    \toprule
    \bfseries ID & \bfseries Model & \bfseries FAD ($\downarrow$) & \bfseries KL ($\downarrow$) &\bfseries CLAP ($\uparrow$)   \\
    \midrule
    0 & 1-rf w/ LN & 1.12 & \bfseries 1.25 & \bfseries 0.659 \\
    2 &1-rf w/o LN & \bfseries 1.08 & 1.27 & 0.649 \\
    1 & 1-rf w/o Init & 1.49 & 1.55 & 0.602 \\
    3 & 1-rf w/o Immi & 1.19 & 1.35 & 0.648 \\
    \midrule
    4 & 2-rf w/ Mix-Exp & \bfseries 1.18 & \bfseries 1.28 & \bfseries 0.658 \\
    5 & 2-rf w/o Mix-Exp & 1.23 & 1.28 & 0.637 \\
    6 & 2-rf w/o Immi  & 1.26 & 1.29 & 0.641 \\
    \bottomrule
    \end{tabular}
    \vspace{1mm}
    \caption{Ablation studies results about FlashAudio. LN: Logit Norm Distribution, Immi: Immiscible Flow, Init: Initialization from pre-trained flow matching. The "w/o LN" and "w/o Mix-Exp" mean the adoption of uniform distribution.}
    \vspace{-5mm}
    \label{table:ablation}
  \end{table}
\section{Conclusion}
In this work, we presented FlashAudio, a novel rectified flow model designed to overcome the limitations of iterative sampling in text-to-audio generation. By introducing Bifocal Samplers and Immiscible Flow, FlashAudio optimized time step allocation and noise distribution, addressing key inefficiencies in the rectified flow process. The proposed Anchored Optimization method further mitigated the impact of classifier-free guidance errors by anchoring the guidance scale to a reference trajectory, thus stabilizing the generation process. Experimental evaluations demonstrated that FlashAudio achieved superior performance in both speed and quality, outperforming existing diffusion-based and consistency-based models, even with one-step generation. Comprehensive analysis and ablation studies confirmed the effectiveness of FlashAudio’s components, highlighting its potential for high-fidelity text-to-audio applications. We envisage that our work could serve as a basis for future text-to-audio studies.
\section{Limitation And Potential Risks}
Despite FlashAudio's notable performance in both few-step and one-step generation settings, it has two primary limitations: (1) GPU resource constraints have limited our ability to explore the scaling potential of our transformer backbone. Future work will focus on expanding the model size of FlashAudio and applying it to a broader range of real-world scenarios. (2) The current model design is restricted to generating fixed-length audio, lacking the capability to produce audio of varying lengths. Addressing these limitations will be a focus of future research efforts.

This paper focuses on achieving efficient and  high-fidelity text-to-audio generation, making generative models more feasible for real-world text-to-audio deployment. Meanwhile it could lead to bias and discrimination. If the training data contains biases, the generative model may not only inherit these biases but also amplify them. This can result in the generated audio content exhibiting discriminatory tendencies related to race, gender, or other social categories.  Such outcomes can be damaging to affected groups and lead to unfair decisions or behaviors, making it crucial to identify and address biases in the training data during model development.

\section*{Acknowledgements}
The research was supported by NSFC (No.62206234) from Mainland China, Early Career Scheme (ECS-HKUST22201322) from Hong Kong RGC, and National Natural Science Foundation of China under Grant No.62222211 and No.U24A20326.

\bibliography{acl_latex}
\clearpage
\appendix
\section{Related Works}\label{A.1}
\subsection{Text-to-audio Generation}\label{A.1.1}
Text-to-Audio Generation is an emerging task that has witnessed notable advancements in recent years. For instance, Diffsound \cite{yangDiffsoundDiscreteDiffusion2023} leverages a pre-trained VQ-VAE \cite{van2017neural} on mel-spectrograms to encode audio into discrete codes, subsequently utilized by a diffusion model for audio synthesis. AudioGen \cite{kreukAudioGenTextuallyGuided2023} frames text-to-audio generation as a conditional language modeling task, while Make-An-Audio \cite{huang2023makeanaudio}, AudioLDM 2 \cite{liuAudioLDMTexttoAudioGeneration2023}, and TANGO \cite{ghosal2023texttoaudio} are all founded on the Latent Diffusion Model (LDM), which significantly enhances sample quality. However, a notable drawback of diffusion models~\citep{liu2023vit,liu2024medic} lies in their iterative sampling process, leading to slow inference and restricting real-world applications. Recent work has thus focused on consistency distillation to achieve few-step and one-step generation. For instance, AudioLCM leverages consistency distillation with a multi-step ODE solver, achieving two-step generation performance comparable to diffusion models. Other works adopt one-step generation but struggle to align with LDMs. In this work, we introduce a novel rectified flow model for fast and high-fidelity text-to-audio generation and achieve superior performance in both few-step and one-step generation.

\subsection{Flow-based Models}\label{A.1.2}
Flow matching~\citep{lipman2022flow} models the vector field of the transport probability path from noise to data samples. Compared to score-based models like DDPM~\citep{ddpm}, flow matching offers more stable and robust training, along with superior performance. Specifically, rectified flow matching~\citep{liu2022flow} learns the transport ODE to follow straight paths between noise and data points as closely as possible, reducing transport costs and enabling fewer sampling steps through the reflow technique. This approach has shown exceptional performance in accelerating image generation~\citep{liu2023instaflow,liu2022rectified,esser2024scaling}.
In the realm of audio generation, Voicebox~\citep{le2024voicebox} employs flow matching to develop a large-scale multi-task speech generation model. Its successor, Audiobox~\citep{vyas2023audioboxunifiedaudiogeneration,ye2024flashspeech}, expands this approach into a unified audio generation model guided by natural language prompts. VoiceFlow~\citep{guo2024voiceflow} introduces rectified flow matching into text-to-speech (TTS), achieving speech generation with fewer inference steps.
However, in the task of text-to-audio (TTA) generation, rectified flow models have not yet been explored to enhance sample efficiency and high-quality audio generation.

\section{Architecture}\label{A.2}
\begin{figure}[htbp]
    \centering
  \includegraphics[width=.49\textwidth]{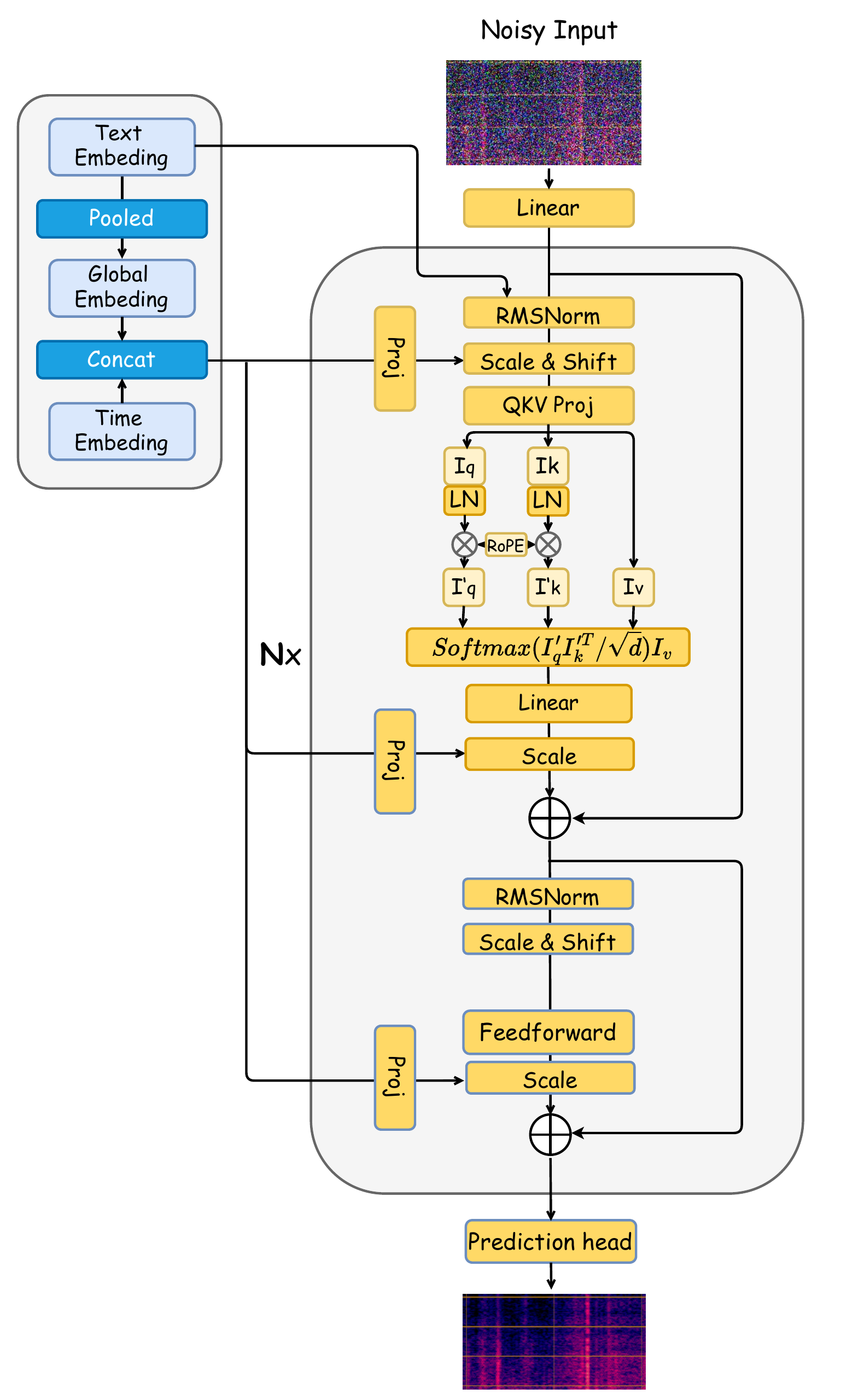}
  \caption{The architecture of FlashAudio.}
  \label{fig:structure}
  \vspace{-4mm}
\end{figure}

We list the hyper-parameters of FlashAudio in Table~\ref{tab:hyper} and the model architecture of FlashAudio is displayed in Figure~\ref{fig:structure}.

\begin{table*}[t]
\small
\centering
\begin{tabular}{l|c|c}
\toprule
\multicolumn{2}{c|}{Hyperparameter}   & FlashAudio \\ 
\midrule
\multirow{5}{*}{Spectrogram Autoencoders} \\
&Input/Output Channels                    & 80      \\
&Hidden Channels                          &   20   \\ 
&Residual Blocks                         &   2   \\   
&Spectrogram Size                &  $80 \times 624$ \\    
&Channel Mult                &   $[1, 2, 4]$ \\   

\midrule
\multirow{6}{*}{Transformer Backbone} \\
&Input shape       &  (20, T)   \\         
&Condition Embed Dim              &  1024 \\ 
&Feed-forward Hidden Size             &  768 \\
&Transformer Heads          &  32 \\
& Transformer Blocks                 &  16\\
\midrule
\multirow{3}{*}{CLAP Text Encoder} \\
&Transformer Embed Channels        &  768   \\         
&Output Project Channels   &  1024 \\    
&Token Length   &  77 \\         
\bottomrule
\end{tabular}
\caption{Hyperparameters of FlashAudio models.}
\label{tab:hyper}
\end{table*}
\section{Additional Quantitative Comparison}\label{A.3}
Due to the page limit, we provide additional quantitative results with performance on text-to-music generation, scalable model size of transformer backbone~\citep{vaswani2023attention,liu2023wav2sql}, and more ablation results on bifocal samplers.
\subsection{Performance on Text-to-Music Generation}\label{A.3.1}
\begin{table*}[ht]
\centering
\small
\begin{tabular}{l|c|cccc|cc}
  \toprule
  \multirow{2}{*}{\bfseries Model} & \multirow{2}{*}{\bfseries NFE} & \multicolumn{4}{c|}{\bfseries Objective Metrics} & \multicolumn{2}{c}{\bfseries Subjective Metrics} \\
                         & & \bfseries FAD ($\downarrow$) & \bfseries KL ($\downarrow$) &\bfseries CLAP ($\uparrow$) & \bfseries RTF ($\downarrow$) &\bfseries MOS-Q($\uparrow$)   & \bfseries MOS-F($\uparrow$)   \\
\midrule
GroundTruth   & / & / & /  & 0.46 & /  &  89.32  & 91.04  \\
Riffusion & /   & 13.31   & 2.10   &  0.19 & 0.40 & 75.23  & 76.12  \\
Mousai & /  & 7.50   & / & /   &  / & / & /  \\
Melody  & /  & 5.41   & / & /   &  / & / & /  \\
MusicLM & /  & 4.00   & / & /   &  / & / & /  \\
MusicGen & / & 4.50  & 1.41   & 0.42 & 1.28  &  81.15  & 84.21  \\
MusicLDM & 200 & 5.20   & 1.47   & 0.40 & 1.40  & 79.34  & 82.10  \\
AudioLDM 2 & 200 &  3.81  & 1.22   &  0.43 & 2.20  &  82.42  & 85.64  \\
AudioLCM & 2 & 3.92 & 1.24 & 0.40 &  0.003 & 82.56 & 85.71 \\
\midrule
\bfseries FlashAudio ($\mathcal{R} $ \usym{2713} $\mathcal{D} $ \usym{2717}) & 4 & \bfseries 3.25 & \bfseries 1.20 &  \bfseries 0.44 & 0.014 & \bfseries 83.96 & \bfseries 85.83 \\
  \bfseries FlashAudio ($\mathcal{R} $ \usym{2713} $\mathcal{D} $ \usym{2713}) & 1 & 3.48 & \bfseries 1.22 & 0.0.42 & \bfseries 0.0025 & 83.14 & 85.06 \\
\bottomrule
\end{tabular}%
\vspace{1mm}
\caption{The comparison between FlashAudio and baseline models on the MusicCaps Evaluation set. We borrow the results of Mousai, Melody, and MusicLM from the MusicGen~\citep{copet2023simple}.}
\vspace{-6mm}
\label{tab:music}%
\end{table*}
We perform a comparative analysis of audio samples generated by FlashAudio against several established music generation systems. These include: 1) GT, the ground-truth audio; 2) MusicGen~\citep{copet2023simple}; 3) MusicLM~\citep{agostinelli2023musiclm}; 4) Mousai~\citep{schneider2023mousai}; 5) Riffusion~\citep{forsgren2022riffusion}; 6) MusicLDM~\citep{chen2024musicldm}; 7) AudioLDM 2~\citep{liu2023audioldm}; 8) AudioLCM~\citep{liu2024audiolcmtexttoaudiogenerationlatent}. The results are presented in Table~\ref{tab:music}, and we have the following observations: (1) In terms of audio quality, FlashAudio outperforms all diffusion-based methods and language models across a spectrum of both objective and subjective metrics with a much less inference time. This highlights FlashAudio's effectiveness in producing high-quality music samples and establishes it as a leading model in audio synthesis. (2) In terms of sampling speed, FlashAudio stands out for its exceptional efficiency. The optimal sampling speed of FlashAudio requires only RTF of 0.0025 while maintaining high-quality output. This illustrates its potent capability to strike an optimal balance between the quality of the samples and the time required for inference.

\subsection{Analyses about Scalable Transformer}\label{A.3.2}
We investigate the performance of a novel transformer backbone designed to scale up the trainable parameters, as showcased in Table~\ref{tab:scalable}. We observe that when the model parameters are reduced to 74M, performance declines across all metrics. However, when the parameter number increases to 429M, there is a performance degradation across multiple metrics. We speculate that this anomalous phenomenon may be due to the convergence difficulty for the larger model under similar training steps, or the redundant model capacity tends to cause overfitting on a relatively small dataset like Audiocaps, deteriorating the model’s generalization performance. Due to the limit of GPU resources, we cannot expand the batch size to a larger value which also has a negative effect on the larger model.

\begin{table}[htbp]
    \centering
    \small
    \vspace{-2mm}
    \begin{tabular}{c|cccc}
        \toprule
        Model & Parameters & FAD & KL & CLAP \\
        \midrule
        Small & 74M & 1.40& 1.43 & 0.639 \\
        Base & 197M & \bfseries 1.18 & \bfseries 1.28 & \bfseries 0.658 \\
        Large & 429M & 1.20 & 1.30 & 0.657 \\
        \midrule
    \end{tabular}
    \caption{The presented figures only account for trainable parameters, i.e., those within the transformer architecture, evaluated on AudioCaps.}
    \vspace{-4mm}
    \label{tab:scalable}
\end{table}

We evaluate the performance of various parameters for the logit-normal and Mix-Exp distributions. The results, summarized in Table~\ref{tab:biofocal}, reveal the following insights: (1) As illustrated in Figure~\ref{fig:sampler}, Logit(-0.5,1.0) and Logit(0.5,1.0) emphasize the left and right regions of the timesteps, respectively, while Logit(0.0,1.0) focuses on the center of the timesteps. The Logit(0.0,1.0) distribution achieves the best performance, which supports our view that 1-RF should prioritize the middle timesteps. (2) Among the Mix-Exp distributions, \(a=4\) produces the best results. Therefore, we adopt \(a=4\) for training 2-RF.

\begin{table}[htbp]
    \centering
    \small
    \vspace{-2mm}
    \begin{tabular}{c|ccc}
        \toprule
        Model  & FAD & KL & CLAP \\
        \midrule
        Logit(-0.5,1.0)  & 1.13 & 1.30 & 0.641 \\
        Logit(0.5,1.0)  & 1.20 & 1.26 & 0.647 \\
        Logit(0.0,1.0) & \bfseries 1.12 & \bfseries 1.25 & \bfseries 0.659 \\
        \midrule
        Mix-Exp(1) & 1.24 & 1.31 & 0.636 \\
        Mix-Exp(2) & 1.23 & 1.33 & 0.638 \\
        Mix-Exp(3) & 1.20 & 1.30 & 0.649 \\
        Mix-Exp(4) & \bfseries 1.18 & \bfseries 1.28 & \bfseries 0.658 \\
        \midrule
    \end{tabular}
    \caption{Performance of different parameters for the logit-normal and Mix-Exp distributions.}
    \vspace{-4mm}
    \label{tab:biofocal}
\end{table}

\section{Evaluation}\label{A.4}
\subsection{Subjective evaluation} \label{A.4.1}
\label{app:human_eval}
To directly reflect the quality of the audio generated, we carry out MOS (Mean Opinion Score) tests. These tests involve scoring two aspects: MOS-Q, which assesses the quality of the audio itself, and MOS-F, which measures the faithfulness of the alignment between the text and the audio.

For assessing audio quality, the evaluators were specifically directed to “concentrate on  quality and naturalness of the audio.” They were provided with audio samples and asked to give their subjective rating (MOS-Q) on a 20-100 Likert scale.

To assess text-audio faithfulness, human evaluators were presented with both the audio and its corresponding caption. They were then asked to answer the question, “Does the natural language description align with the audio?” The raters had to select one of the options: “completely,” “mostly,” or “somewhat,” using a 20-100 Likert scale for their response.

Our crowd-sourced subjective evaluation tests were conducted via Amazon Mechanical Turk where participants were paid \$8 hourly.

\subsection{Objective evaluation} \label{A.4.2} 
The Fréchet Audio Distance (FAD)~\cite{kilgour2018fad}, adapted from the Fréchet Inception Distance (FID) for the audio domain, is a reference-free perceptual metric designed to measure the difference between the distributions of generated audio and ground truth audio. FAD is commonly used to assess the quality of generated audio content.

KL divergence is calculated at the level of paired samples between the generated audio and the ground truth audio. This metric is determined by the label distribution and is then averaged to produce the final result.

CLAP score, adapted from the CLIP score~\cite{hessel2021clipscore,radford2021learning} for the audio domain, is a reference-free evaluation metric to measure audio-text alignment for this work that closely correlates with human perception.


\section{Bifocal Samplers for Rectified Flow Models}\label{A.5}
\begin{figure}[b]
  \includegraphics[width=.46\textwidth]{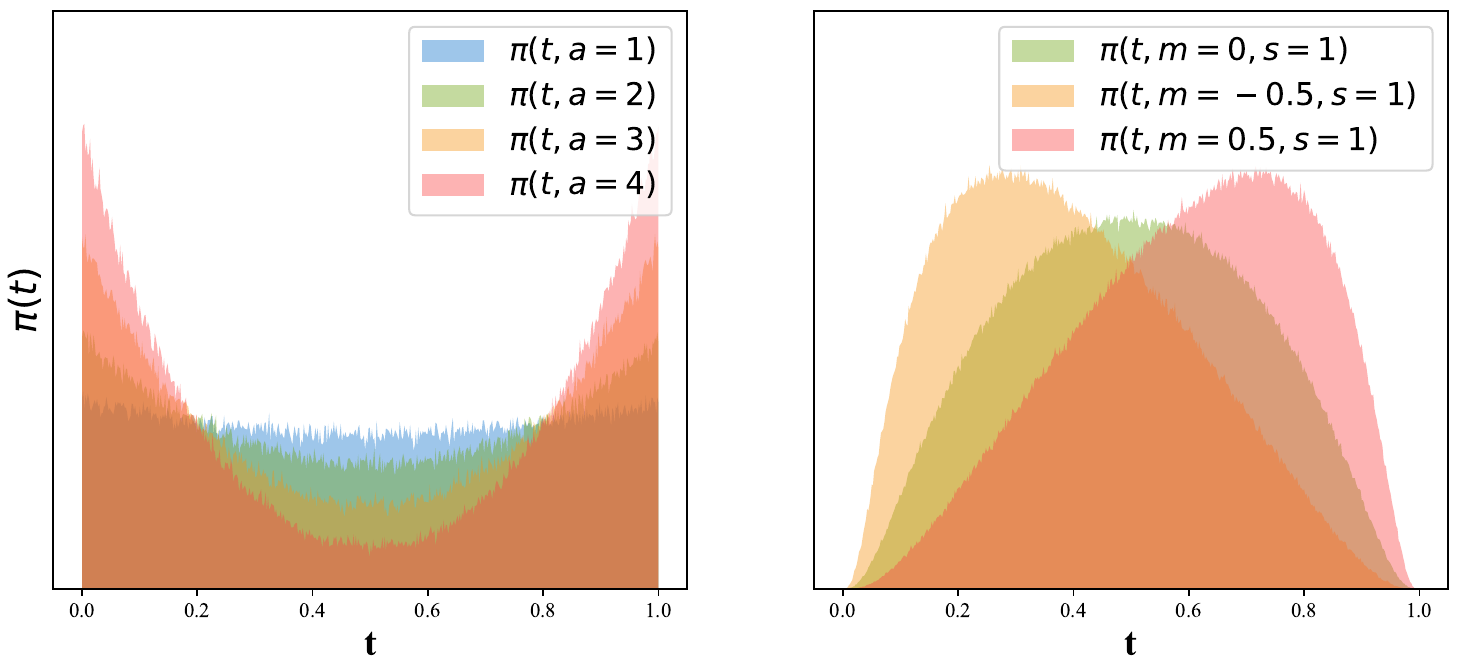}
  \caption{The Mix-Exp distribution (left) and logit-normal distributions (right) ,which are used in  biasing the sampling of training timesteps.}
  \label{fig:sampler}
  \vspace{-4mm}
\end{figure}
As discussed in Section\ref{Bifocal} we introduce a novel sampler named Bifocal sampler which combines a mixture of exponential distribution (Mix-Exp) and logit-normal distribution in different processing stages respectively ,which leads to obvious improvement. The distribution of both Mix-Exp sampler and logit-normal sampler are been visualized in Figure~\ref{fig:sampler}. The peaks of these two distributions are located at the middle of t and the boundary of t, respectively. The use of two distinct peaks at different stages is why we refer to this sampling strategy as the Bifocal Samplers.

\begin{figure*}[htbp]
	\centering
    \subfigure[One-step Generation Comparisons]
    {
    \includegraphics[width=0.95\textwidth]{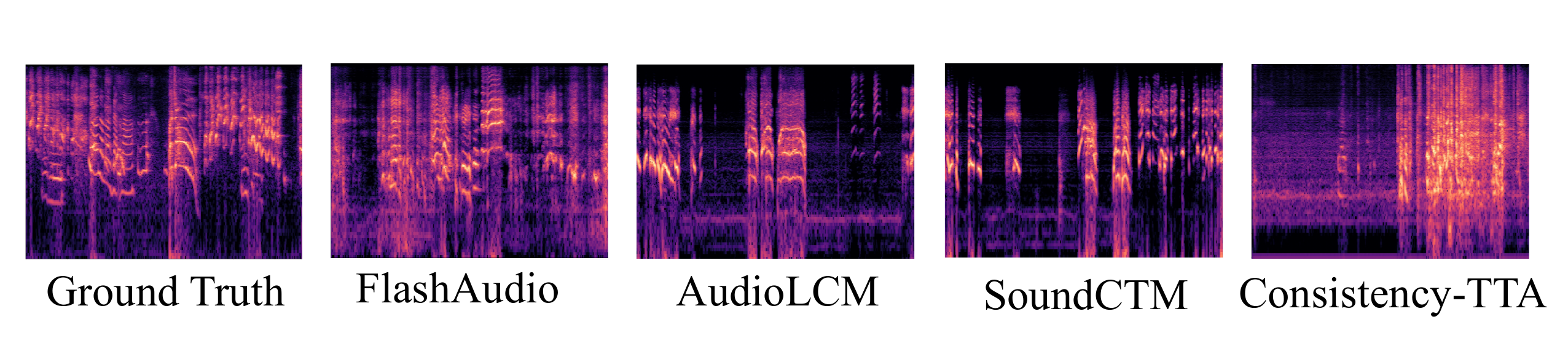}
    }
    \subfigure[Multi-step Generation Comparisons]
    {
    \includegraphics[width=0.95\textwidth]{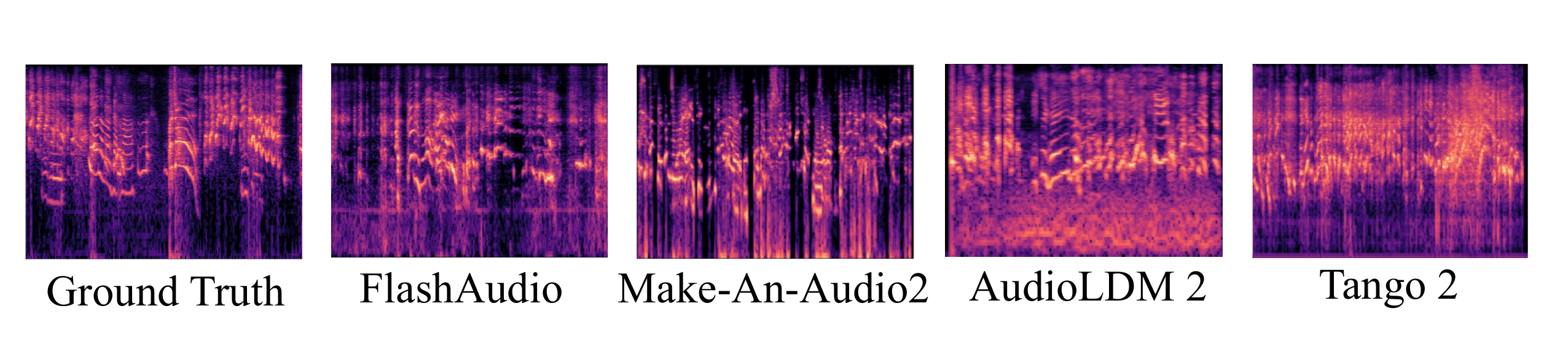}
    }
	\caption{Mel-Spectrums for caption: A man and woman laughing followed by a man shouting then a woman laughing as a child laughs.}
	\label{fig:mels}
\end{figure*}

\section{QUALITATIVE RESULTS} 
We visualize FlashAudio with mel-spectrograms and compare it with baselines in both multi-step generation and one-step generation setting in Figure~\ref{fig:mels}. FlashAudio is prone to generate the most similar mel-spectrum with the ground-truth while other baselines like Consistency-TTA produces noisy output and SoundCTM deviate from the GT. Compared to diffusion-based models, FlashAudio shows the capabilities to generate clear and high-fidelity mel-spectrum.



\begin{figure*}[htbp]
	\centering
    \subfigure[Screenshot of MOS-F testing.]
    {
    \includegraphics[width=0.95\textwidth]{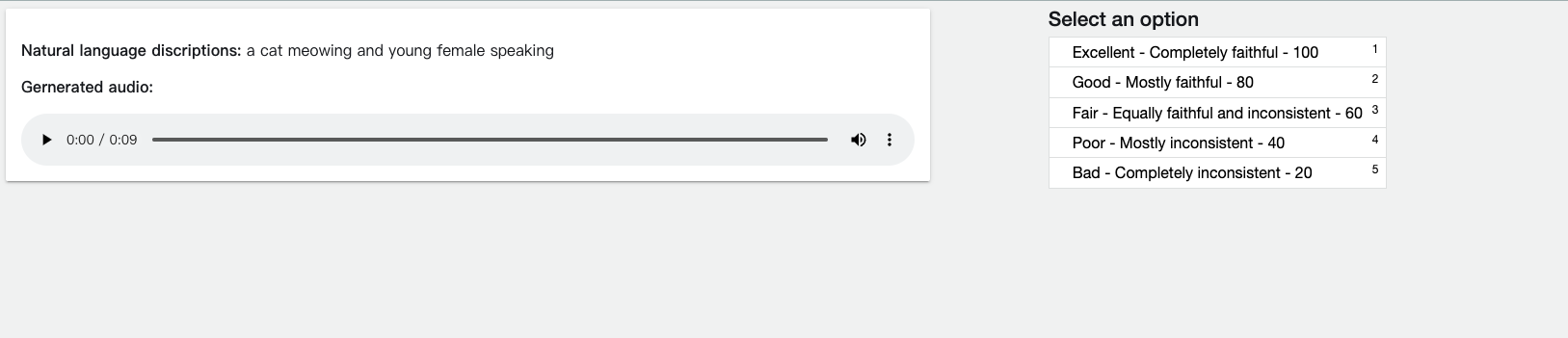}
    }
    \subfigure[Screenshot of MOS-Q testing.]
    {
    \includegraphics[width=0.95\textwidth]{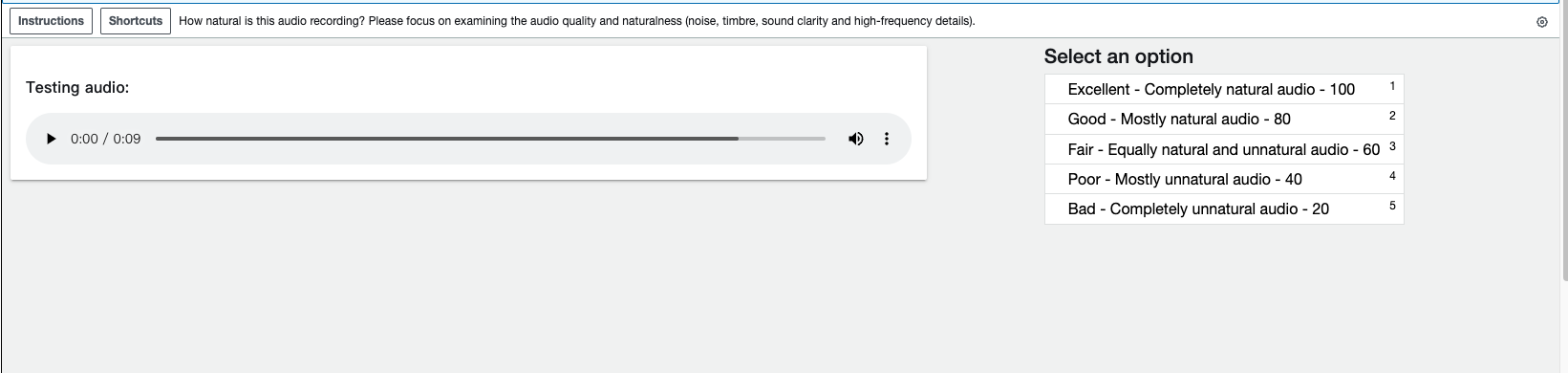}
    }
	\caption{Screenshots of subjective evaluations.}
	\label{fig:screenshot_eval}
\end{figure*}

\end{document}